\documentstyle[prl,aps]{revtex}
\input{psfig}
\draft
\begin{document}
\twocolumn[\hsize\textwidth\columnwidth\hsize\csname@twocolumnfalse\endcsname
\title{Controlled Dynamics of Interfaces in a Vibrated Granular Layer}
\author{I. Aranson, D. Blair,  W. Kwok, G. Karapetrov, U. Welp, 
G. W. Crabtree,   V.M.  Vinokur}  
\address{
Argonne National Laboratory,
9700 South  Cass Avenue, Argonne, IL 60439} 
\author{L.  Tsimring}
\address{Institute for Nonlinear Science, University of California, San Diego, La Jolla,
CA 92093-0402
}
\date{\today}
\maketitle

\begin{abstract} 
We present  experimental study of 
a topological excitation, {\it interface}, in 
a vertically vibrated layer of granular material. We show that these
interfaces, separating regions of granular material 
oscillation with  opposite phases, can be shifted 
and  
controlled by a very small amount 
of an additional subharmonic signal, mixed with the harmonic 
driving signal. 
The speed and the  direction of interface motion depends sensitively 
on the phase and the amplitude 
of the subharmonic driving. 
\end{abstract} 


\narrowtext
\vskip1pc]

Despite their ubiquity and many practical applications,
an understanding of the fundamental dynamical behavior of
granular materials  remains a serious challenge\cite{jnb}.
One of the main obstacles  for the development of
a continuous description of granular flow is  the difficulty in
performing 
quantitative experiments under
controlled conditions. Testing of theoretical models of
the basic excitations of the system is 
especially important. 
It has been shown
recently\cite{swin1,swin2,swin3,jaeger1} that  thin
layers of granular materials subjected to vertical
vibration exhibit a diversity of patterns which may play the role of
such fundamental excitations.  
The particular pattern is determined by the interplay between driving frequency
$f$ and the acceleration amplitude $\Gamma= 4 \pi^2 {\cal A }  f^2/g$ of the 
cell, 
where  
${\cal A} $ is the amplitude of oscillation and 
$g$ is the acceleration due to gravity. 
Periodic patterns, such as squares and   stripes,  or localized 
oscillons vibrating with frequency $f/2$   appear  at  
 $\Gamma\approx 2.4$ \cite{swin1,swin2,swin3,jaeger1}.  
At higher acceleration
($\Gamma>3.72$), stripes and squares become unstable, and hexagons appear
instead. Further increase of acceleration
replaces hexagons with a non-periodic structure of interfaces 
separating large domains of flat layers oscillating 
with opposite phase
with frequency $f/2$. 
These interfaces were called kinks in 
Ref. \cite
{swin2,swin3,blair}. These interfaces
are either smooth or "decorated'' by
periodic undulations depending on parameters \cite{swin3,blair}.
For $\Gamma >5.7$,  various
quarter-harmonic patterns emerge.
Several theoretical approaches  
including molecular dynamics simulations, order parameter equations and 
hydrodynamic-type models, 
have been proposed  recently to describe this phenomenology, see e.g. \cite{at,theor}. 

In this Letter we present an experimental study of the dynamics of 
interfaces in a vibrated thin layer of granular material. 
We find that an additional 
subharmonic driving results in a controlled displacement of the 
interface. In the absence of  subharmonic driving, the interface drifts 
toward the middle of the cell. 
When the subharmonic frequency $f/2$ is slightly detuned, 
the interface moves periodically about the
middle of the cell.  The present  experimental results are in agreement with 
theoretical predictions \cite{atv}. 

Interfaces in a granular layer separate regions of granular material
oscillating with opposite phases with respect to the bottom plate of the 
vibrating cell. 
These two  phases are  related to the
period-doubling character of the flat layer motion at large plate 
acceleration. 
Since an interface separates two stable symmetric dynamic phases, 
it can be interpreted as  a topological 
defect, similar to a domain wall in ferromagnets separating regions of 
opposite magnetization \cite{ferro}.   
Interfaces 
can only disappear at the walls of the cell or
annihilate with other interfaces.
The existence of the interfaces can be understood
from the following consideration.
Since grains in a layer lose their kinetic 
energy in multiple inter-collisions during landing at the plate,  
the behavior of a granular layer can be compared 
with the dynamics of a fully inelastic ball bouncing on a plate 
vibrating  with amplitude ${\cal A}$ and frequency $f$ \cite{swin2,behr2}.  
In this case, for driving with acceleration
$\Gamma $ less than $\Gamma_0
\approx3.72$  
the ball lifts to the same height at each  cycle \cite{swin2,metha}. 
Above $\Gamma_0$ the motion  exhibits period-doubling, i.e. the heights of 
elevation alternate at each cycle. As a result, for 
$\Gamma > \Gamma_0$, the two states of the bouncing ball,
differing by the initial phase,  would coexist. If the analogous 
states of the bouncing flat layer  
co-exist in different parts of the cell, they have to be separated by 
an interface. These interfaces, found  experimentally 
\cite{swin2,swin3,blair} and theoretically \cite{atv}, are flat 
for high frequency drives and show  transverse instability
leading to periodic decoration at lower frequencies (see Fig. 1). 
In an infinite system a straight interface 
is  immobile due to the symmetry between alternating states: 
the motion of flat layers on both sides of the interface 
is identical with a phase shift $\pi$. 
Additional driving at the subharmonic frequency $f_1=f/2$  will break 
the symmetry between domains with opposite phases. Depending on the 
phase of the additional driving, $\Phi$, with respect to the phase of the 
primary driving, the relative velocity
of the layer and the plate at collision will differ 
on different sides of the interface, and the material on one side  
will be lifted to a larger height than on the other side. 
As a result, the direction of the interface motion
can be controlled by the phase $\Phi$. The speed at a given 
$\Phi$ is determined by the amplitude of 
the subharmonic acceleration $\gamma$. 

In our previous  work we have developed a phenomenological description of
the pattern formation in thin layers of granular material \cite{at,atv}. 
On the basis of our order parameter model  
 we have predicted\cite{atv} that for small values of 
$\gamma$,  the interface velocity $V= dX/dt$, $X$ being  the position
of the interface, is a periodic function  of  $\Phi$,
\begin{equation} 
\frac {d X}{ d t } = V_0   \sin(\Phi - \Phi_0)
\label{eq1} 
\end{equation} 
where $V_0 = \alpha \gamma$, and susceptibility
$\alpha$ and phase shift $\Phi_0$ are  functions of 
the driving amplitude $A$ and frequency $f$ which can be 
estimated using the model Ref. \CITE{atv}. 

We performed experiments with 
a thin layer of granular material subjected to two-frequency driving. 
Our experimental setup 
is similar to that of  \cite{swin1,swin2,swin3}. 
We used  $0.15$mm diameter bronze balls, and the layer thickness 
in most of our experiments was 10 particles. We performed 
experiments in a circular cell of 15.3 cm  diameter and in 
a rectangular cell of $4\times12$ cm$^2$, which were evacuated to 
2 mTorr. We varied the
acceleration  $\Gamma$ and the frequency $f$ of the primary driving signal 
as well as 
acceleration $\gamma$, frequency $f_1$ and the  phase $\Phi$ of the 
secondary (additional) 
driving signal. The interface  position and vertical accelerations  
were acquired  using a CCD  camera and accelerometers and further 
analyzed on a Pentium computer.
The lock-in technique was used 
to measure  the accelerations at $f$ and $f/2$ frequencies 
in real time.

In the absence of additional 
driving the interface drifts toward the middle of the cell
(see Fig. 1). We attribute this  effect to the feedback between the 
oscillating granular layer and  the plate vibrations due to 
the finite ratio of the mass of the granular material to the 
mass of the vibrating plate. 
Even in the absence of subharmonic drive, the vibrating cell 
can acquire a subharmonic motion from the periodic impacts of the 
granular layer on the bottom plate at half the driving frequency. 
If the interface is located in the middle of the cell, 
the masses of material on both sides of the interface are equal, 
and due to the anti-phase character of the layer motion on both
sides, an additional subharmonic driving force is not produced. 
The displacement of the interface $X$ from the center of cell leads to a mass
difference $\Delta m$ on opposite sides of the interface which in turn causes 
an additional subharmonic driving proportional to $\Delta m$. 
In a rectangular cell $\Delta m \propto  X$.
Our experiments show that
the interface moves in such a way to decrease the subharmonic response,
and  the feedback provides an additional term $-X/\tau$ in the r.h.s. of
Eq. (\ref{eq1}), yielding 
\begin{equation}
\frac {d X}{ d t } = -  X/\tau +  V_0 \sin(\Phi - \Phi_0)
\label{eq2}
\end{equation}
The relaxation time constant $\tau$ depends on the mass ratio  (this also
holds for the circular cell if $X$ is small compared to the cell radius).
Thus, in the absence of  an additional
subharmonic drive ($\gamma=0$), 
the interface will eventually divide the cell into
two regions of equal area (see Fig. 1). 

In order to verify this model we performed the 
following experiment. We positioned the interface 
off center by applying 
an additional subharmonic drive. Then we turned off the 
subharmonic drive and immediately measured  
the amplitude $\mu$ of the plate acceleration at the subharmonic 
frequency\cite{note}.  The results are presented in Fig. 2. 
The subharmonic acceleration of the 
cell decreases exponentially as the interface propagates to the 
center of the cell. 
The measured relaxation time $\tau$ of the subharmonic 
acceleration  decreases
with the mass ratio of the granular layer and of the cell with 
all other parameters fixed. 
The mass of the granular layer was varied by 
using two different cell sizes  while keeping the thickness of the layer 
unchanged.
For the cells shown 
in Fig. 1, we found that the relaxation time $\tau$ 
in the rectangular cell is about 4 times 
greater than for the circular cell (see Fig. 2). This  is 
consistent with the ratio of the total masses of granular material
(52 grams in rectangular cell and 198 grams in circular one). 
In a separate experiment 
the mass of the cell was changed by attaching an additional weight 
of 250 grams  to the 
moving  shaft, which weights 2300 grams. 
This led to an increase   
of the  corresponding relaxation times of 15-25 \%. 
The relaxation time $\tau$  increases rapidly with $\Gamma$ 
(see Fig. 2, inset). 

When an additional subharmonic driving is applied, 
the interface is displaced from 
the middle of the cell. 
For small amplitude of the subharmonic 
driving $\gamma$,  the stationary interface position is $X=  V_0 
 \tau  \sin(\Phi - \Phi_0) $, 
since the restoring  force balances the external driving force.
Fig. 3 shows the positions of the interface 
for various  $\Phi$. From such images we determined 
the equilibrium position $X$ as function of $\Phi$ (Fig. 4a). 
The solid line depicts the sinusoidal fit predicted by the theory. 
Because of the finite mass ratio effect, the amplitude of the 
measured plate acceleration $\mu$ at 
frequency $f_1$ also shows periodic behavior with 
$\Phi$, (see Fig. 4b), enabling us to  infer the
interface displacement from the acceleration measurements. 
For even larger amplitude of subharmonic driving (more than 
4-5 \% 
of the primary driving) extended patterns (hexagons) re-emerge 
throughout the cell. 

The velocity $V_0$ which the interface would have in an infinite system, 
can be found from 
the measurements of the relaxation time $\tau$ and maximum displacement 
$X_m$ at a given amplitude of subharmonic acceleration $\gamma$,
$V_0=X_m/\tau$, see Eq.(\ref{eq2}). 
We verified that in the rectangular cell 
the displacement $X$ depends linearly on $\gamma$ almost up to values 
at which the interface disappears at the short side wall 
of the cell (see Fig. 5, inset). 
Figure 5 shows the susceptibility  $\alpha=V_0/\gamma$ as a function of the
amplitude of the primary acceleration $\Gamma$. The susceptibility decreases 
with $\Gamma$.  The cusp-like features in the $\Gamma$-dependence of $\alpha$
(and $\tau$, see inset to Fig. 2), 
are presumably 
related to the commensurability between the lateral size of the cell and the
wavelength of the interface decorations.

We developed an alternative 
experimental technique which allowed us to measure 
simultaneously the relaxation time $\tau$ and the ``asymptotic'' 
velocity $V_0$.  This was achieved 
by a small detuning $\Delta f$ of the additional 
frequency $f_1$ from the 
exact subharmonic frequency $f/2$, i.e. $\Delta f = f_1 -f/2 \ll f$. 
It is equivalent to the linear increase of phase shift $\Phi$ with the 
rate $2 \pi  \Delta f $. This linear growth of the phase results in 
a periodic motion of the interface with frequency $\Delta f $ and
amplitude $X_m=V_0/\sqrt { \tau^{-2} +
(2 \pi \Delta f)^2}$ (see Eq (\ref{eq2})). 
The measurements of the ``response functions" 
$X_m(\Delta f)$ are presented in Fig. 6. 
From the  dependence of $X_m$ on $\Delta f$  we can extract 
parameters $V_0, \alpha$ and $\tau$ by a fit  to the 
theoretical function.  
The measurements are in very good agreement with 
previous independent measurements of 
relaxation time $\tau$ and susceptibility $\alpha$. 
For comparison with the previous results, we indicate
the values for $\tau$ and $\alpha$,
obtained from the response function measurements of Figs. 2 and 5 
(stars). The measurements agree within   $5$ \%.

\begin{figure}
\leftline{\psfig{figure=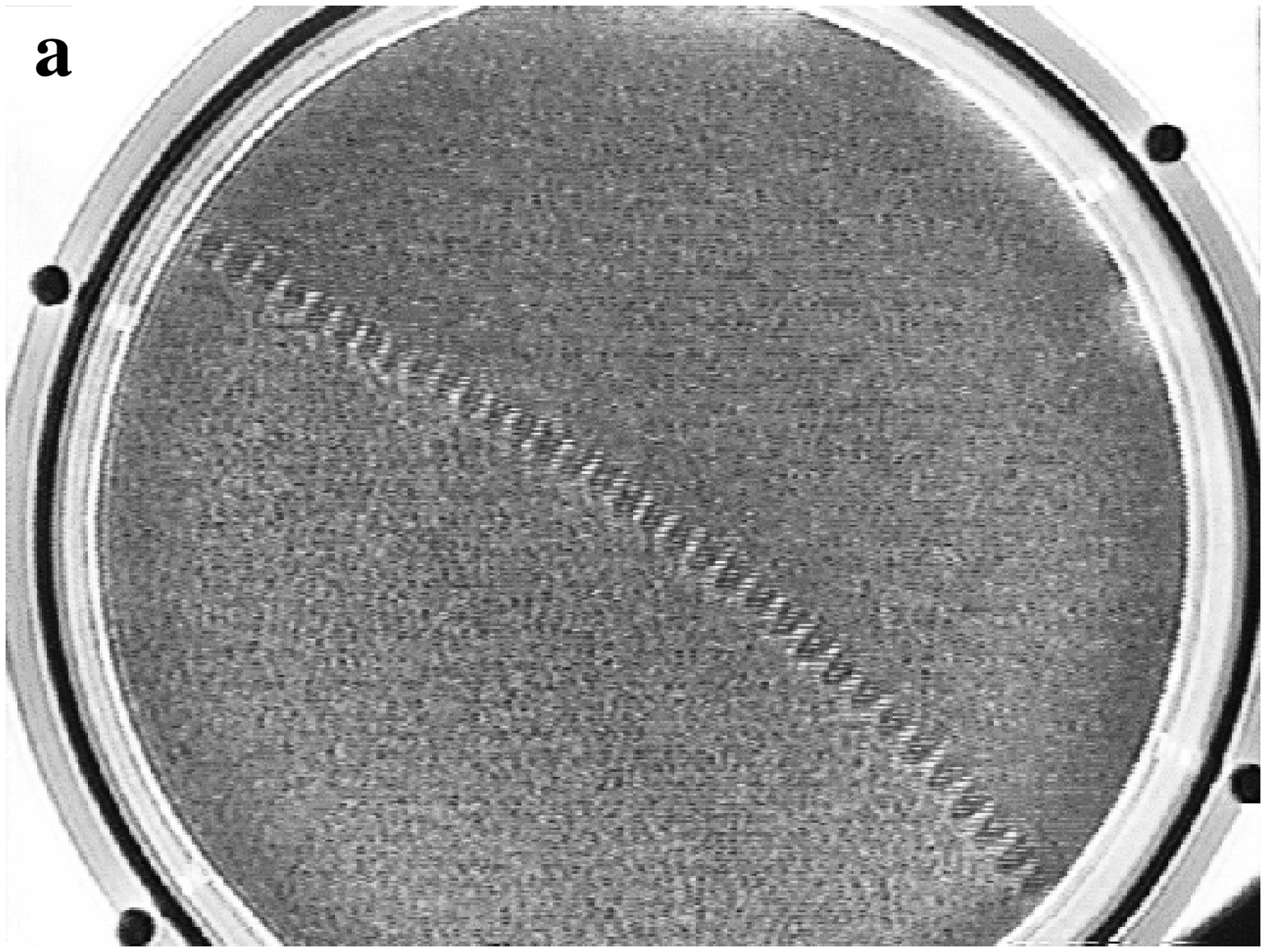,height=2.5in}}
\leftline{\psfig{figure=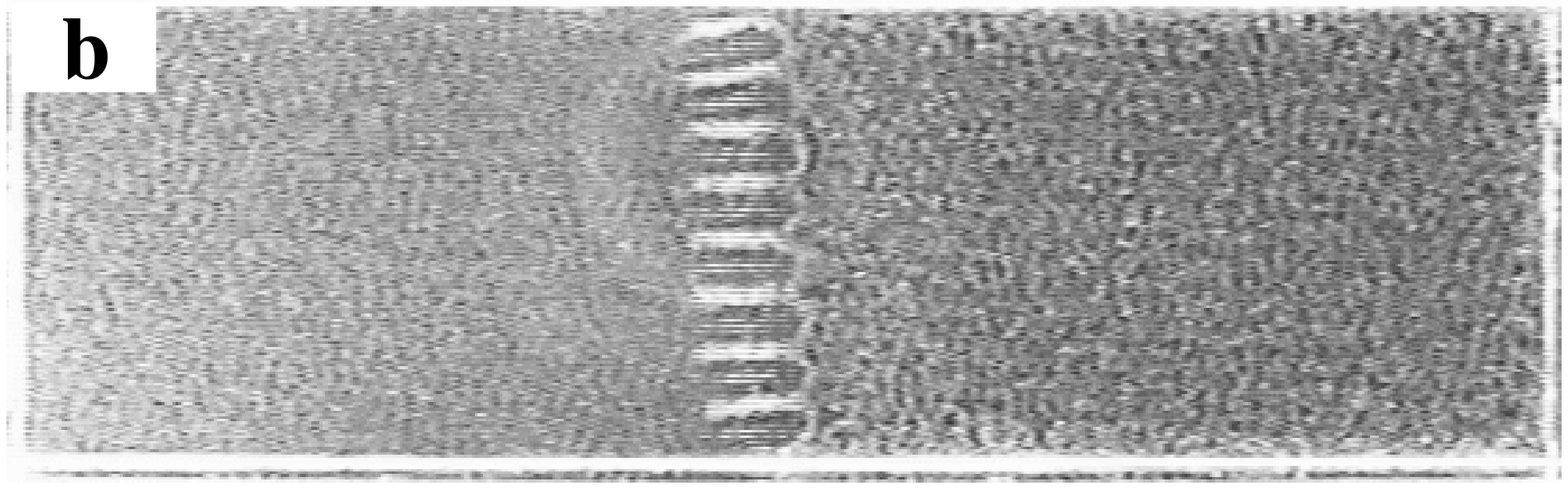,height=1.05 in}}
\caption{Stationary position of the interface in circular, diameter $15.3$ cm
 (a) and rectangular $4\times 12 $ cm (b) cells
for  sinusoidal driving force
with  $f=40$ Hz and  
 $\Gamma=4.1 $. Layer thickness 10 particle diameters.}
\end{figure}

In summary, the  position of a vertically vibrated
granular layer can be  controlled  by a
very small acceleration applied at the subharmonic frequency
(of the order of 0.1\% of the primary harmonic acceleration).
The direction and magnitude of the interface displacement depend
sensitively on the relative phase of the subharmonic drive.
Our measurements confirm the theoretical predictions made
on the basis of the order parameter model\cite{atv}.
We observed that period-doubling motion of the flat layers
produces subharmonic driving because of the finite ratio
of the mass of the granular layer and the cell. This in
turn leads to the restoring force driving the interface towards the middle
of the cell.

We thank L. Kadanoff, R.  Behringer,  H. Jaeger,  H. Swinney
and P. Umbanhowar
for useful discussions. This research is supported by
US DOE, grants \#  W-31-109-ENG-38, DE-FG03-95ER14516,  and by NSF,
STCS \#DMR91-20000.

\begin{figure}
\leftline{\psfig{figure=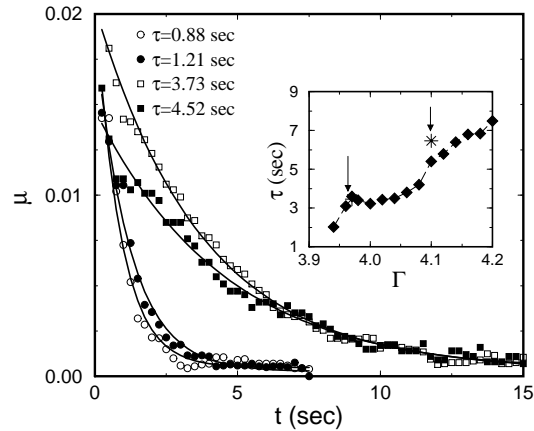,height=2.5in}}
\caption{Amplitude of the  subharmonic acceleration
$\mu $ of  the cell averaged over 4 measurements vs 
time for the interface propagating
to the center of the cell for $\Gamma = 3.97$ and
$f=40$ Hz.
Circles/squares correspond to the circular/rectangular
cells,  open/closed symbols correspond to
light/heavy cells, respectively. Heavy cells differ from
light ones by an additional weight of 250 g
attached to the moving shaft.
Solid lines show exponential fit
$\mu \sim  \exp (t/\tau ) + const $. 
Inset: $\tau$ vs $\Gamma$ for light rectangular cell.}
\end{figure}

\begin{figure}
\leftline{\psfig{figure=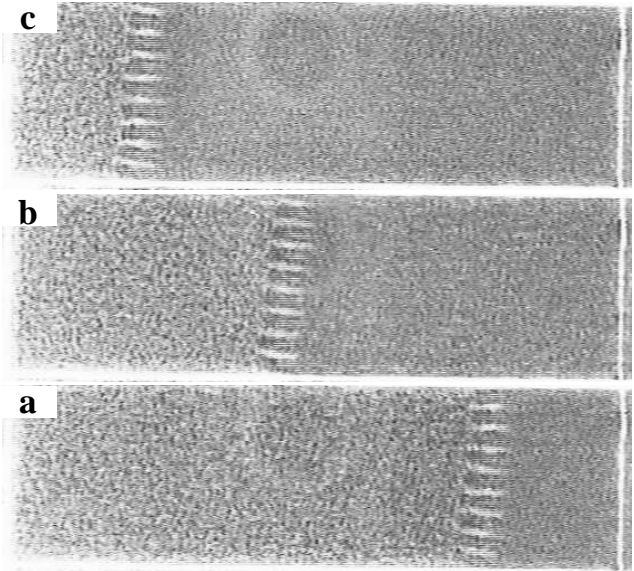,height=3. in}}
\caption{Equilibrium position of 
interface for $\Phi=80^0$ (a); $\Phi=170^0$ (b); $\Phi=260^0$ (c)
for $\Gamma=4.1$, $f=40$Hz, $f_1=f/2$, and $\gamma=0.6 \%$ of 
$\Gamma$ in a  rectangular cell. 
 }
\end{figure}

\begin{figure}
\leftline{\psfig{figure=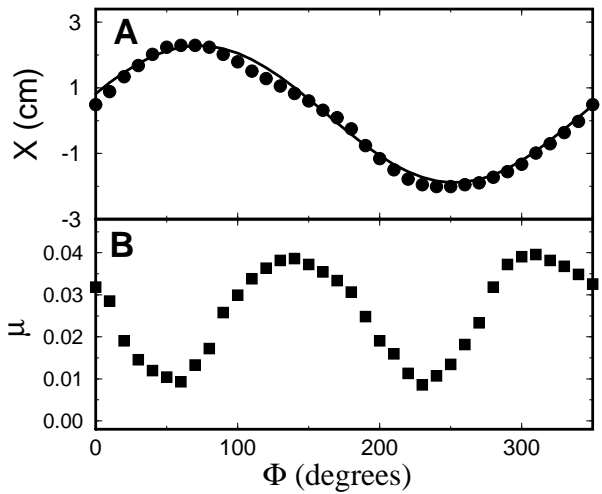,height=2.9 in}}
\caption{
(a) 
Equilibrium position $X$ (b) amplitude of measured 
subharmonic acceleration $\mu$   as functions of phase $\Phi$.  
Circular cell, 
$\Gamma=4.1$, $f=40$ Hz, $\gamma=1.25 \% $ of $\Gamma$. 
}
\end{figure} 

\begin{figure}
\leftline{\psfig{figure=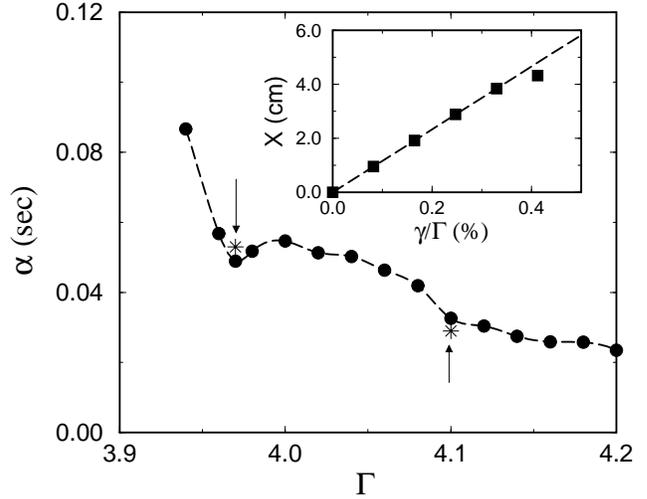,height=2.9 in}}
\caption{
Susceptibility $\alpha=V_0/\gamma$ vs $\Gamma$ 
at $f=40$Hz,  rectangular cell. 
Inset: Displacement $X$ as function of $\gamma$ at $\Phi=260^0$. 
}
\end{figure} 

\begin{figure}
\leftline{\psfig{figure=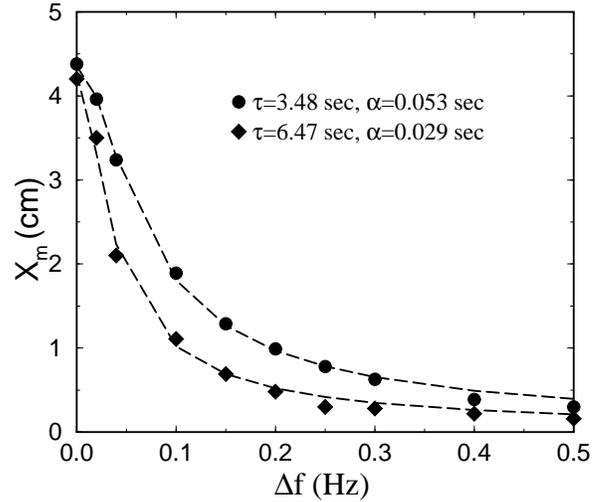,height=2.9 in}}
\caption{
Maximum displacement $X_m$ from center of rectangular 
cell as function of 
frequency difference $\Delta f = f_1 -f/2$
for $f=40$ Hz, and 
for $\Gamma=3.97$ (circles) and $\Gamma=4.1$ (diamonds).
Dashed lines
are  fit to $X_m = V_0/\sqrt{\tau ^{-2} + 
(2 \pi \Delta f)^2}$. 
The values of $\alpha$ and $\tau$ obtained from the fit are also indicated in 
Figs. 2 and 5 (stars).
}
\end{figure}
\end{document}